\documentclass[manuscript]{aastex}
\usepackage{pdflscape}
\usepackage{multicol} 
\usepackage{epstopdf}
\usepackage{graphicx}
\usepackage{color}

\shorttitle{Application of the Trend Filtering Algorithm for Photometric Time Series Data}
\shortauthors{Gopalan ET AL.}

\begin{document}

\title{Application of the Trend Filtering Algorithm for Photometric Time Series Data}

\author{Giri Gopalan\altaffilmark{1}, Peter Plavchan\altaffilmark{2}, Julian van Eyken\altaffilmark{3}, David Ciardi\altaffilmark{3}, Kaspar von Braun\altaffilmark{4}, Stephen R. Kane\altaffilmark{5}}

\altaffiltext{1} {Harvard Medical School, supported by a 2009 Caltech Summer Undergraduate Research Fellowship; Giridhar\_Gopalan@hms.harvard.edu. Although employed by HMS this research is not in any way associated with his employment.}
\altaffiltext{2}{Department of Physics, Astronomy and Materials Science, Missouri State University, 901 S National Ave., Springfield, MO 65897}
\altaffiltext{3}{NASA Exoplanet Science Institute, California Institute of Technology, M/C 100-22, 770 South Wilson Avenue, Pasadena, CA 91125; plavchan@ipac.caltech.edu}
\altaffiltext{5}{Lowell observatory visiting astronomer}
\altaffiltext{5}{Department of Physics and Astronomy, San Francisco State University}

\begin{abstract}
Detecting transient light curves (e.g., transiting planets) requires high precision data, and thus it is important to effectively filter systematic trends affecting ground based wide field surveys. We apply an implementation of the Trend Filtering Algorithm (TFA) \citep{kovacs1} to the 2MASS calibration catalog and select Palomar Transient Factory (PTF) photometric time series data.  TFA is successful at reducing the overall dispersion of light curves, however it may over filter intrinsic variables and increase ``instantaneous'' dispersion when a template set is not judiciously chosen. In an attempt to rectify these issues we modify the original literature TFA by including measurement uncertainties in its computation, including ancillary data correlated with noise, and algorithmically selecting a template set using clustering algorithms as suggested by various authors. This approach may be particularly useful for appropriately accounting for variable photometric precision surveys and/or combined data-sets. In summary, our contributions are to provide a MATLAB software implementation of TFA and a number of modifications tested on synthetics and real data, summarize the performance of TFA and various modifications on real ground based data sets (2MASS and PTF), and assess the efficacy of TFA and modifications using synthetic light curve tests consisting of transiting and sinusoidal variables. While the transiting variables test indicates that these modifications confer no advantage to transit detection, the sinusoidal variables test indicates potential improvements in detection accuracy.
\end{abstract}


\section{Introduction}

Recent technological advancements in astronomy, including the ability to store massive amounts of observational data, have allowed astronomers to explore the time domain in detail. These advancements have allowed for the production of  detailed photometric time series data, or ``light curves'', which monitor the brightness of a stellar object on time-scales varying from minutes to decades.  The analysis of light curves has led to a prodigious number of exoplanet discoveries in recent times; according to the NASA Exoplanet Archive, to date over 1262 exoplanets have been detected by analyzing photometric time series data to look for transits. 

Photometric time series data obtained from wide field surveys are affected by systematic noise from varying atmospheric conditions or uncorrected instrumental effects. The ability to intelligently filter out such systematic noise is crucial to detecting transiting planets or variable stars, as shown by \citet{hartman}, and a number of methods have been developed to mitigate their spurious effects, including regularized regression, GP-regression, principal components analysis, and unsupervised learning methods \citep{Wang, kim}. While these methods have been developed primarily in the context of Kepler, our focus is that of ground based wide field surveys, where atmospheric conditions play a dominant role in contributing to unwanted systematic noise. The Trend Filtering Algorithm (TFA; Kovacs et al. 2005) attempts to detrend systematic noise in light curves, and it leverages the fact that wide field surveys generate multiple light curves that are affected by similar systematics. Our overall research efforts focused primarily on: i) implementing this algorithm on existing photometric time series data sets from ground based wide field astronomical surveys and ii) investigating methods to improve its performance. We note that TFA can also be used in the signal reconstruction phase, as is investigated by \citet{aigrain} and \citet{fore}, but do not focus on this version of TFA for the purposes of our analysis.

In ${\S}$ 2, we briefly discuss the 2MASS and PTF Orion pilot data we have applied TFA to. In ${\S}$ 3, we analyze in detail the original version of TFA as well as various modifications to TFA we have implemented. In ${\S}$ 4, we discuss the results of performing TFA on simulated data sets with varying template selection procedures and two types of light curves: those with uniform errors, and those with measurement uncertainties that vary nightly, and two types of signals: transits and sinusoids. We also present the results of the detrending on the aforementioned data sets. In ${\S}$ 5 we discuss these results and conclude. 

\section{Data Selection}
\subsection{2MASS Data}

The Two Micron All Sky Survey (2MASS) is a wide field survey which captured all portions of the sky in three bands of the near infrared spectrum over a period of four years, from 1997 to 2001. For each night of 2MASS operations, all of the survey telescopes were targeted at one of 35 calibration fields every hour using the  scanning strategy used for the main survey \citep{plavchan2}. Over the entirety of the survey, between 562 and 3692 observations were made of each of the calibration fields. This raw data was preprocessed using the system used to process the survey observation data \citep{skrutskie},  which detected and extracted source positions and photometry for all objects in the images from each scan. The standard stars in each field were used to determine the nightly photometric zero-point solutions for each time point, in addition to seasonal atmospheric parameters. All data extracted from these scans were stored in the 2MASS Calibration Point Source Working Database (Cal-PSWDB), which contains more than 191 million source extractions from 73,230 scans of the 35 calibration fields \citep{cutri}. 

\subsection{PTF Orion Pilot Data}
Data for the Palomar Transient Factory (PTF) Orion project were obtained from 40 nights of observation in December 2009 to January 2010 using the PTF camera installed at the Palomar 48 inch telescope. The goal of the observations was to study the young 25 Ori association \citep{julian}. For our analyses,  we used preliminary pilot test data taken during instrument commissioning time in Feb/March 09, with the purpose of using them to build a differential photometry pipeline. Two data sets were taken. The first was an overlapped tiling of the entire Orion region, R band, 30s exposures. The second was a single-field time-series data set, R band, 60s exposures, approximately 90s cadence, over three nights. The aim of the program was to observe a field where the gas disks of young stars are actively on the point of dissipating (approximately 5--10Myr old), leaving behind any newly formed Jovian planets. The 25 Orion region was chosen for the time-series pilot study.

\section{Analysis}

In ${\S}$ 3.1, we begin by briefly summarizing the methodology of TFA \citep{kovacs1} as well as the derivation of the matrix formulation of the algorithm. In the remaining sections, we analyze various methods of modifying TFA to improve its performance. In ${\S}$ 3.2 we discuss how one may include measurement uncertainties and recast this modification into matrix algebra. In ${\S}$ 3.3, we review two clustering based approaches to optimize the selection of a template set, the first method from \citet{kim}. In the final section we present a method of including external parameters correlated with noise known as External Parameter Decorrelation, as in \citet{bakos}.  

\subsection{Formulation of TFA}
We begin with the mathematical formulation of TFA \citet{kovacs1}. Let $Y(i)$ be a light curve which is to be detrended, and assume it is zero averaged.  TFA assumes that a filter function $F(i)$ may be constructed as a linear combination of ``template'' light curves $X_j(i)$ which are selected from the field surveyed, and are zero averaged as well. Implicitly, the template set contains all information of the systematic trends that the algorithm is privy to.  Assume there exist $m$ light curves in the template set, and assume each light curve consists of $n$ brightness measurements; to summarize we have the following relations:
\begin{eqnarray}
F(i) &=&  \displaystyle\sum_{j = 1}^{j = m}c_jX_j(i) \\
Y^{*}(i)&=& Y(i)-F(i)
\end{eqnarray}
where $Y^{*}(i)$ is the filtered light curve. Thus the formulation of a filter function is equivalent to the solution of a particular set of constants $c_j$. TFA solves for these constants by minimizing the following sum of squared residuals:
\begin{eqnarray}
min\displaystyle\sum_{i = 1}^{i = n}(Y(i)-F(i))^2
\end{eqnarray}
This minimization problem can be recast in terms of matrix algebra; we may arrange the template set into an $m$ by $n$ matrix $L$ where each row is a light curve. Further, consider placing the constants $\{c_j\}$ into a length $m$ row vector. Then we have the relation:
\begin{eqnarray}
F = cL
\end{eqnarray}
Where $F$ is the filter function viewed as a length $n$ row vector. Using this new notation, our minimization problem now becomes:
\begin{eqnarray}
min ||Y-cL||_2
\end{eqnarray}
Where the $||.||_2$ notation is the familiar notion of Euclidean distance in $\mathbb{R}^n$. In other words, a particular choice of $c$ corresponds to a point in the subspace spanned by the rows of $L$, and we seek a choice of $c$ such that the residual distance $Y-cL$ is minimized. A basic proposition from linear algebra dictates that this choice of $c$ will cause the vector $Y-cL$ to be normal to the hyperplane spanned by the rowspace of $L$. Hence we seek  $c$ such that the vector $Y-cL$ is normal to all the rows of $L$. Since normality is defined by a 0 inner product, we get the following relation:
\begin{eqnarray}
L(Y-cL)^T = 0
\end{eqnarray}
Hence solving for $c$:
\begin{eqnarray}
c = (LL^T)^{-1}LY^T
\end{eqnarray}
The salient point of this derivation is that a filter function can be calculated via basic matrix operations, making for a simple \texttt{MATLAB} implementation.

\subsection{Including Measurement Uncertainties}
A main drawback of TFA is that is does not rely on individual measurement uncertainties, and hence a highly uncertain measurement is treated equally to a very certain measurement. To rectify this, one may modify the key minimization problem TFA employs to weight certain measurements more than uncertain measurements. Specifically, we now minimize the following weighted sum of squared residuals:
\begin{eqnarray}
min\displaystyle\sum_{i = 1}^{i = n}w_i(Y(i)-F(i))^2
\end{eqnarray}
where $w_i = \sigma_i^{-2}$.  As before, we may recast this problem in terms of matrices. If we define the diagonal matrix $S$ such that $S_{ii} = \sigma_i^{-2}$ where $\sigma_i$ is the standard error of the $i_{th}$ brightness measurement, then we are minimizing the following:
\begin{eqnarray}
min ||(Y-cL)S||_2
\end{eqnarray}
Since $S$ is a linear map this is equivalent to the problem:
\begin{eqnarray}
min ||(YS-cLS)||_2
\end{eqnarray}
This is of the same form as the original TFA problem and hence the solution is given by:
\begin{eqnarray}
c = (GG^T)^{-1}GH^T
\end{eqnarray}
Where
\begin{eqnarray}
G = LS 
\end{eqnarray}
and
\begin{eqnarray}
H = YS 
\end{eqnarray}
It should be noted that the involvement of measurement uncertainties is more computationally costly than the unmodified version of TFA. With the original algorithm, one may form a template set for an entire field of stars and and use the same template for each star filtered. However, if one corrects by measurement uncertainties the template set must be multiplied by $S$ for each star filtered, where $S$ is dependent on the particular star. Weighting by measurement uncertainties is suggested in the thesis of \citet{andras}.

 Another variation to consider is to choose the weights to be a function of template measurement uncertainties in addition to uncertainties of the light curve being filtered. This is a variation we do not consider in our current analysis but could be interesting to investigate in future work. Nonetheless when the clustering procedure described in the next section is used, one ultimately averages over light curves in a cluster to produce an element of the template set. This should reduce the measurement standard error associated with the template light curves approximately on the order of square root of $N_C$ (assuming independent and identical measurement uncertainty), where $N_C$ is the number of light curves used in a particular cluster; hence the measurement uncertainty of the light curve to be filtered will be of more importance when the clustering procedure is used to  produce templates.

\subsection{Optimizing the Template Set}
A critical component of TFA is the selection of a template set, since it implicitly contains all information about systematic noise. Ideally, one would like to minimize the number of template stars while maximizing information about systematic noise. While a large number of template stars yields a large reduction in dispersion, this also yields a greater tendency to over filter intrinsic variables. This is because many free parameters in the minimization problem allows for many degrees of freedom to fit a particular light curve and potentially filter intrinsic variations. We investigate two methods of optimizing the selection of a template set, both based on clustering algorithms.
\subsubsection{Agglomerative Hierarchical Clustering}
\citet{kim} propose an algorithm which attempts to select a small number of template stars that well represent systematic trends. The algorithm is, in essence, an implementation of agglomerative hierarchical clustering. The algorithm aims to partition stars into subsets whose light curves correlate highly with each other. The logic behind this approach is that each cluster represents a particular sort of systematic trend. Once partitioning is complete, one may extract a template star from each cluster by performing a weighted average, where weighting is done by the inverse of variance.  
\newline\indent
There are three main steps involved in this algorithm. First a distance matrix is computed for the light curves. Then, one computes a binary tree using this distance matrix. Finally, one uses the binary tree to determine clusters via a merging algorithm. We detail these steps below:

\begin{itemize}
\item \textbf{STEP 1}: \textit{Compute the Distance Matrix.} First, we compute the Pearson correlation between all light curves. We store the information in a distance matrix $D$ where $d_{ij} = 1-c_{ij}$, where $c_{ij}$ is the correlation between light curves $i$ and $j$.
\item \textbf{STEP 2}: \textit{Compute the Binary Tree.} We then compute a binary tree using the distance matrix. Specifically, the leaves of the tree are the individual light curves. We then combine the closest two nodes under one parent, where the distance between two nodes $a$ and $b$ is the maximum distance between any two light curves in the nodes. We iterate this linking procedure until all light curves have been merged.
\item \textbf{STEP 3}: \textit{Determine Clusters via Merging.} Using the binary tree, one can determine clusters in the following manner. Initially we set clusters to be nodes in the tree with at least two children, and at each step consider merging the two closest nodes to form a larger cluster. Call this potential cluster \textit{Cmerge}; if  \textit{Cmerge} contains light curves which are correlated (i.e \textit{Cmerge} is a good representation of a particular systematic trend) then the distribution of distances between any two light curves in \textit{Cmerge} should (approximately) follow a normal distribution. Hence, one applies an Anderson-Darling normality test to the list of distances. If the test produces a p-level below .10 (i.e, we have reason to believe the distances do not come from a normal distribution) then we stop the merging procedure, as we have evidence that the light curves are not all correlated with each other. In this fashion, one may partition all light curves into subsets which are all correlated which each other. Once the clusters are formed, one takes a weighted average of light curves in the cluster to produce a template trend. 
\end{itemize}

\subsubsection{KMEANS Clustering}
\indent An alternate approach to clustering is the KMEANS algorithm. If one assumes that all light curves are elements of $\mathbb{R}^n$ where $n$ is the number of brightness measurement for each light curve, then one may formulate a notion of Euclidean distance between two light curves.  Using this notion of distance, KMEANS partitions a set of light curves into $k$ subsets where the elements of each subset are close to each other. One begins the algorithm by assigning $k$ random points in $\mathbb{R}^n$ as centers. Next, one assigns each light curve to the center it is closest to, and in this process we partition the set into $k$ subsets. Then, we recalculate $k$ centers by choosing the average of each cluster as its center. We then iterate this process of reassigning light curves to clusters and recalculating centers until no new assignments have been made. A subtlety involved is the initial choice of centers, and a particularly efficient method for doing so is given by KMEANS++ \citep{arthur}. Note that one must ensure light curves are zero averaged to ensure that Euclidean distance is a feasible metric. An additional subtlety in KMEANS is that the $k$ must be chosen beforehand, in contrast to the hierarchical clustering approach. 

\subsection{External Parameter Decorrelation}
While TFA assumes one has no apriori information regarding systematic noise, it is feasible that certain external parameters, such as seeing or position, correlate with noise. \citet{bakos}  suggest a method to involve external parameters which correlate with systematics. In essence, the formulation is the same as TFA, except coefficients are now chosen for the parameters via the same minimization problem. 
\begin{eqnarray}
min\displaystyle\sum_{i = 1}^{i = n}w_i(Y(i)-\sum_{j=1}^{j = m}c_jX_j(i) -\sum_{j=m+1}^{j=l}c_jP_j(i))^2
\end{eqnarray}
Where $P_{m+1}$ through $P_{l}$ contain the external parameters. Again, we may recast this formulation into matrix algebra; we simply add additional rows to the template matrix $T$, where each row contains the external parameter values for each time index.

\section{Results}
We have written a package of MATLAB software which implements the Trend Filtering Algorithm as well as the aforementioned extensions.\footnote{This is freely available at \texttt{https://github.com/ggopalan/MATLAB-Detrending-Software}} What follows are a series of quantitative assessments of the algorithm, using both the unmodified version and the various modifications previously reviewed. 

\subsection{Simulation Studies}
Before presenting the results of both the modified and unmodified versions of TFA applied to the aforementioned data sets, we discuss two simulation studies conducted to determine the efficacy of the methodology discussed in the previous section. The first simulation study included injected transit signals and the second injected pulsating signals.
\subsubsection{Injected Transit Study} 
This simulation study consists of two sets of 432 light curves with faint transit signals of randomly varying period length, depth and duration added to a single sinusoidal systematic. The transit period was drawn uniformly within $\pm$ 10 $\%$ of 1,4,20,and 400 days, the depth was either .005, .01, or .02 mag, and the transit duration fraction was either .005, .002, or .01. \footnote{The transit duration fraction is how long the transit lasts as a fraction of the orbital period.} Furthermore,  each systematic had a period 1,4,20 or 400 days exactly and 3 possible amplitudes of .005, .01 or .02 mag. The first of these sets also includes fixed measurement errors and the second consists of measurement errors which fluctuate on a nightly basis, but are fixed throughout a given night for all light curves; the periods of the transit signal and systematic were selected to be on similar scales to allow for a particularly challenging recovery scenario. The uniform measurement standard deviation is .01 mag, and the nightly varying measurement standard deviation is uniformly sampled from the range .005 mag to .015 mag for a given night. Ultimately the measurement error is drawn from a  zero mean Gaussian distribution with the associated standard deviation. In total there were 19200 measurements with 1 measurement per minute with 8 hours of observations over 40 nights. This particular cadence was chosen to be representative of a typical ground-based transit survey strategy. These key parameters are summarized in Table 1.

We next considered 3 template sets to filter these same sets of light curves; the first was a template set generated by randomly selecting 25 percent (108) of the 432 light curves for the analysis, the second was chosen with agglomerative hierarchical clustering (the PDT function of our MATLAB package) and the third template set was derived by randomly selecting a set of the same cardinality as that from the agglomerative hierarchical approach (which were 18 and 15 light curves for the uniform and nightly error synthetics). We ran the unmodified and modified TFA (with measurement uncertainties) using these three template sets, which resulted in 6 total filtering runs. To detect whether the resultant light curves had a transit signal we used the BLS periodogram tool \citep{kovacs3} to determine if the correct transit period was detected to a statistically significant extent, where statistical significance is determined by a p-value  $\leq .01$. To detect a signal we used the highest peak if it was statistically significant and considered a maximum of 10 integer multiple subharmonics for signal detection. 

The details of the transit detection methodology used are as follows. We search periods between 0.5 and 40 days with periods evenly spaced in frequency.  We employ the highest peak + correct frequency criterion, and include up to $M/N$ integer fraction multiples (e.g., harmonics) of the correct frequency with $M,N \in (1,2,3,4)$.  For a peak period $P_1$ and the true period $Q$, we define the correct frequency score as: $Max(10^{(-10000*M*N*(Log(P_1)/Log(Q*M/N) -1)^2)})$ for $M,N \in (1,2,3,4)$.  The 10,000*M*N constant determines the width of range of acceptable periods about the true period or its harmonic, with a penalty for higher harmonics.  We also assess the statistical significance of the highest peak utilizing the lognormal distribution of periodogram power values \citep{akeson}.  Secondary peaks that were statistically significant were also investigated, but not included since often only the highest peak is considered in typical ground-based transit surveys.

The results of this simulation study are summarized in Table 2. The results of the synthetics test indicate that TFA with the largest number of template trends results in the largest number of detections. However, in both data sets TFA with the agglomerative hierarchical clustering approach detects more transits than TFA without clustering. Overall the detection rate of transits after applying TFA with a (relatively) small template set yields a comparable detection rate to that of TFA applied with a (relatively) large number of templates; it must be noted that the detection rate of transits is strictly smaller for the modified TFA using clustering than the unmodified TFA with a randomly selected large set of templates.

\subsubsection{Injected Sinusoid Study}
We had conducted a second simulation study with sinusoidal signals instead of transits using the Lomb-Scargle periodogram \citep{scargle, zech} instead of the BLS algorithm for transits.   
As in the previous simulation study, there was an observation every minute for 8 hours over 40 nights. Each systematic had a period 1,4,20 or 400 days exactly and 3 possible amplitudes of .005, .01 or .02 mag. Each true sinusoidal signal had 4 possible periods selected uniformly at random within +/- 10 percent of 1, 4, 20, or 400 days and exactly 3 possible amplitudes of .005, .01, or .02 mag. Table 3 displays the results of this test, where as before, 432 lightcurves were generated with a single systematic and signal, and a signal was considered recovered if the period was correctly recovered with a p-value $\leq .01$. In all cases except the  Pre-TFA uniform light curve set, the modified version of TFA using clustering and uncertainty weighting recovers the most number of signals, although our conclusions are limited by the generally small number of recovered signals.

\subsection{Assessing the Unmodified Version of TFA on Real Data}
\subsubsection{TFA Reduces Dispersion}
By visualizing dispersion versus apparent magnitude for both the 2MASS and PTF Orion pilot data, we have determined that TFA reduces the overall dispersion of the light curves. These graphs are depicted below in \textbf{Figure 1}. Specifically, the algorithm reduced the dispersion most substantially for the PTF Orion pilot data.  

\subsubsection{TFA May Filter Intrinsic Variables}
Evidence suggests that TFA may over filter intrinsic variable signals if many template light curves are chosen. For example, consider \textbf{Figure 2}, which presents the original and filtered light curve for a PTF variable candidate with a template set of 50 light curves that are arbitrarily chosen. The amplitude of the variability of the raw light curve of this source is too large (>0.15 mag) to be explained by systematic sources of photometric variability, from the instrument/telescope, and/or the Earth's atmosphere.  It is also an outlier in its photometric rms for its apparent magnitude, which also allows us to flag the variations we see as being intrinsic to the source rather than systematic.  In this example it is clear that any intrinsic variability has been flattened by the filtering algorithm. This light curve also demonstrates that TFA may be increasing its ``instantaneous dispersion''. In other words, while the overall standard deviation may be reduced, the standard deviation for points taken within a small time frame has increased, which is not a desired result.

\subsection{Assessing the Modified Version of TFA on Real Data}

\subsubsection{Modified TFA Reduces Dispersion}
It is evident that the modifications made to TFA do not mitigate its ability to reduce the dispersion of light curves; the dispersion plots presented in \textbf{Figure 1} are qualitatively similar to dispersion plots of the unmodified version of TFA. In addition to these dispersion plots, we have generated histograms which illustrate the distribution of dispersion reductions. In particular, \textbf{Figure 3} present histograms of dispersion improvements. These histograms illustrate both absolute and relative dispersion improvements. Absolute improvement is defined as the difference between the dispersion of light curves post and pre TFA. Relative dispersion improvement is defined as the absolute improvement divided by the pre TFA dispersion. From these histograms, we can derive  quantitative measure of the modified TFA's performance. On average, dispersion was reduced by 30 percent for the PTF data set and 1.5 percent for the 2MASS data set. The lack of significant improvement to most 2MASS light curves indicates that the data reduction for 2MASS and calibration is thorough in removing most systematics. However, significant improvements can be obtained in special cases, such as extended or confused sources.

\subsubsection{Modified TFA May Prevent  the Overfiltering of Intrinsic Variables}
In some cases, our modified TFA no longer overfilters intrinsic variables.For example, consider the detrended PTF Orion pilot light curve  in \textbf{Figure 2} which was previously over filtered by the unmodified version of TFA. The intrinsically variable source is essentially unchanged. In addition, the problem of increased instantaneous dispersion is also alleviated, although it still persists to a slight extent. This success can be most likely attributed to the prudent selection of a few template trends. The utilization of many template curves allows for many free parameters in the key minimization problem that TFA employs; in turn, this allows intrinsic variations to be filtered by the template curves. 

\section{Conclusion}
After implementing TFA and applying it to the 2MASS calibration data and PTF Orion pilot data we conclude that the algorithm substantially reduces the dispersion of light curves, most notably for the PTF Orion pilot data set. Based on real examples from the PTF Orion pilot data and the synthetic pusating variables test, it is also apparent that TFA may over filter intrinsic variables and increase the instantaneous dispersion of light curves when a template set is not carefully chosen. By modifying TFA to include measurement uncertainties, include ancillary data correlated with noise, and select a template set using clustering algorithms, we believe that these effects can be mitigated. Additionally, we believe our implementation is equipped to handle night-to-night and airmass dependent variable photometric precision for the fixed integration time ground-based surveys due to the involvement of measurement uncertainties. Finally, we note that the MATLAB package and documentation is freely available at \textit{https://github.com/ggopalan/MATLAB-Detrending-Software}. 

\section{Acknowledgements}
GG would like to acknowledge the gracious support of the Caltech Summer Undergraduate Research Fellowship program for supporting this work during the summer of 2009.

\begin{figure}
\plottwo{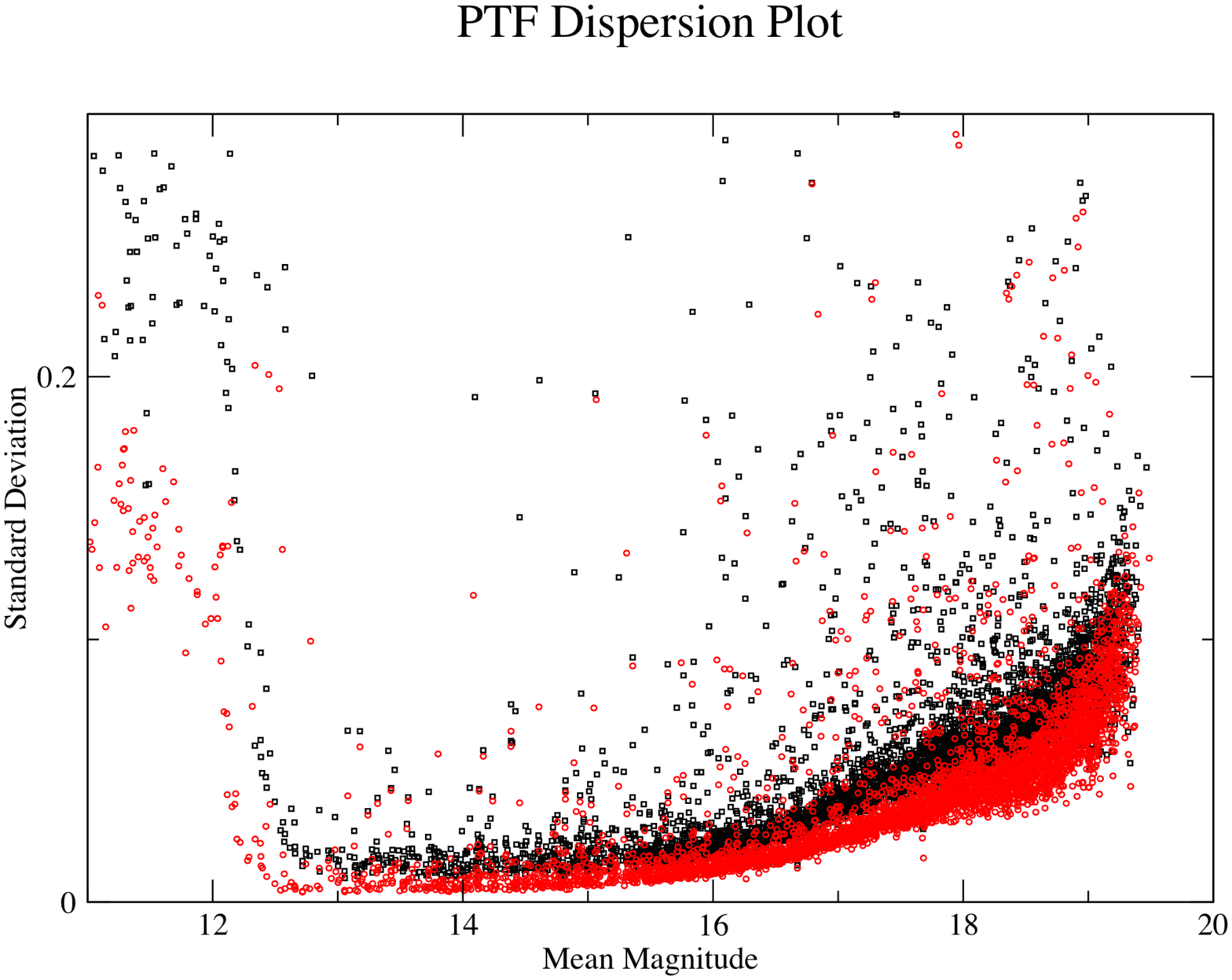}{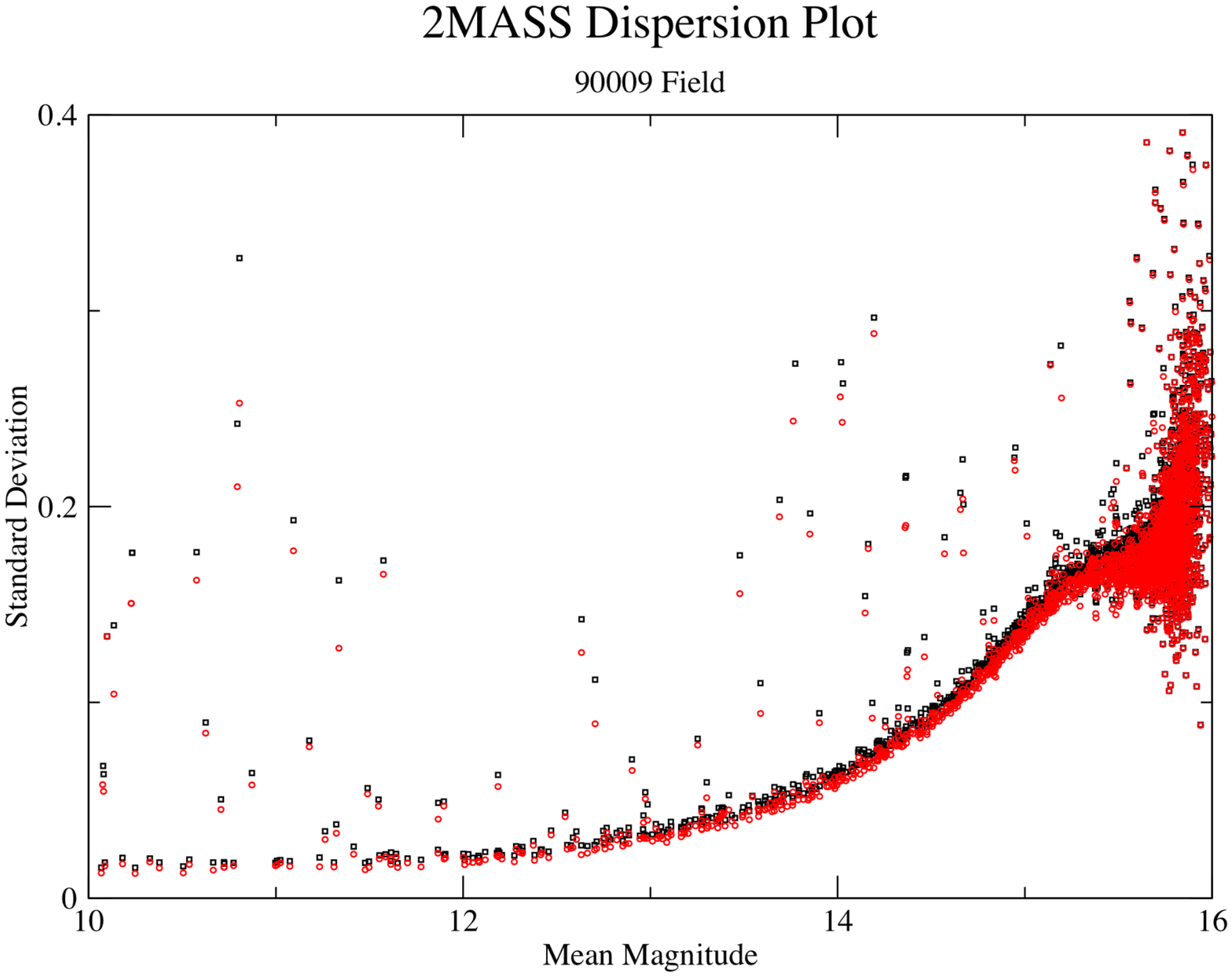}
\plottwo{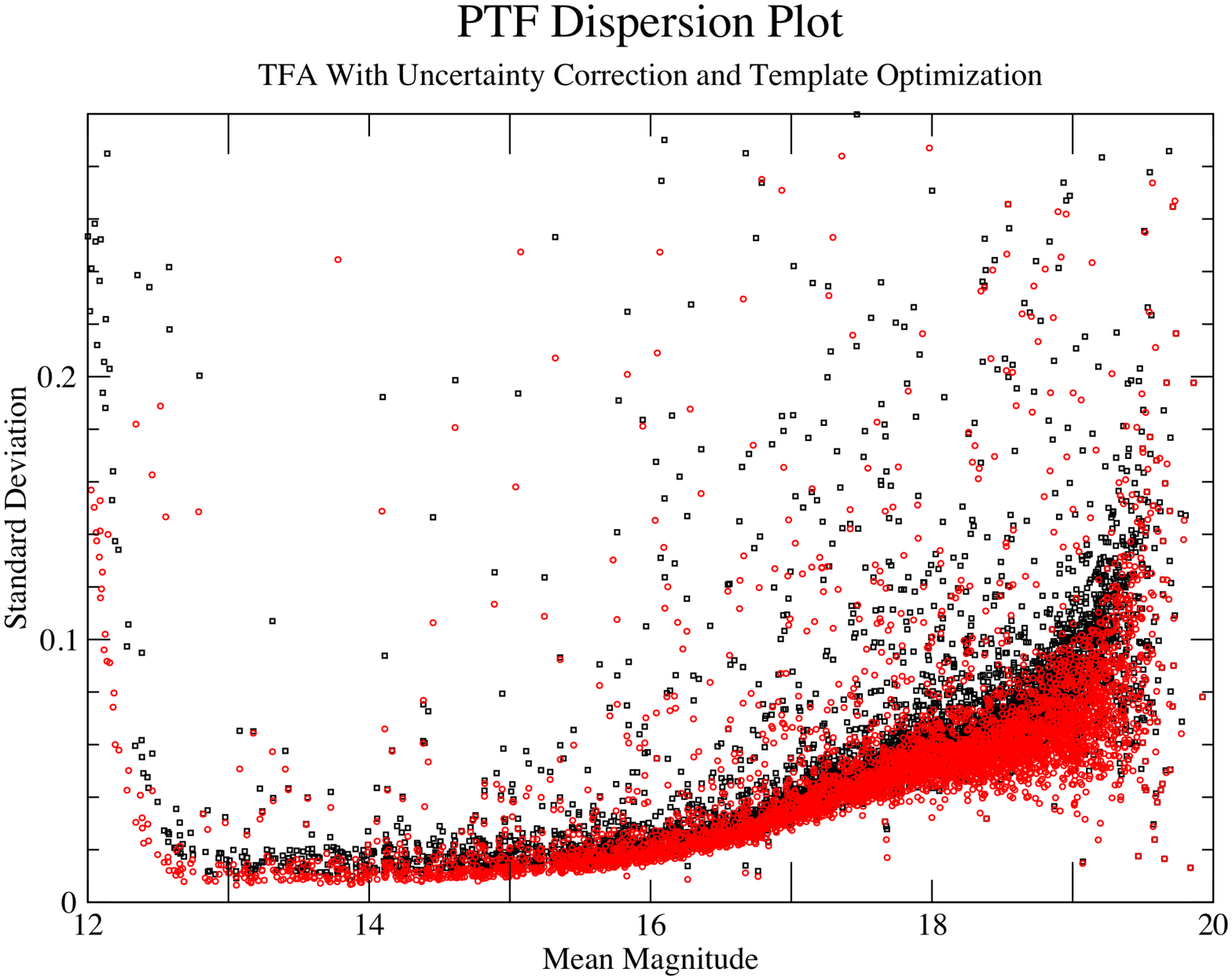}{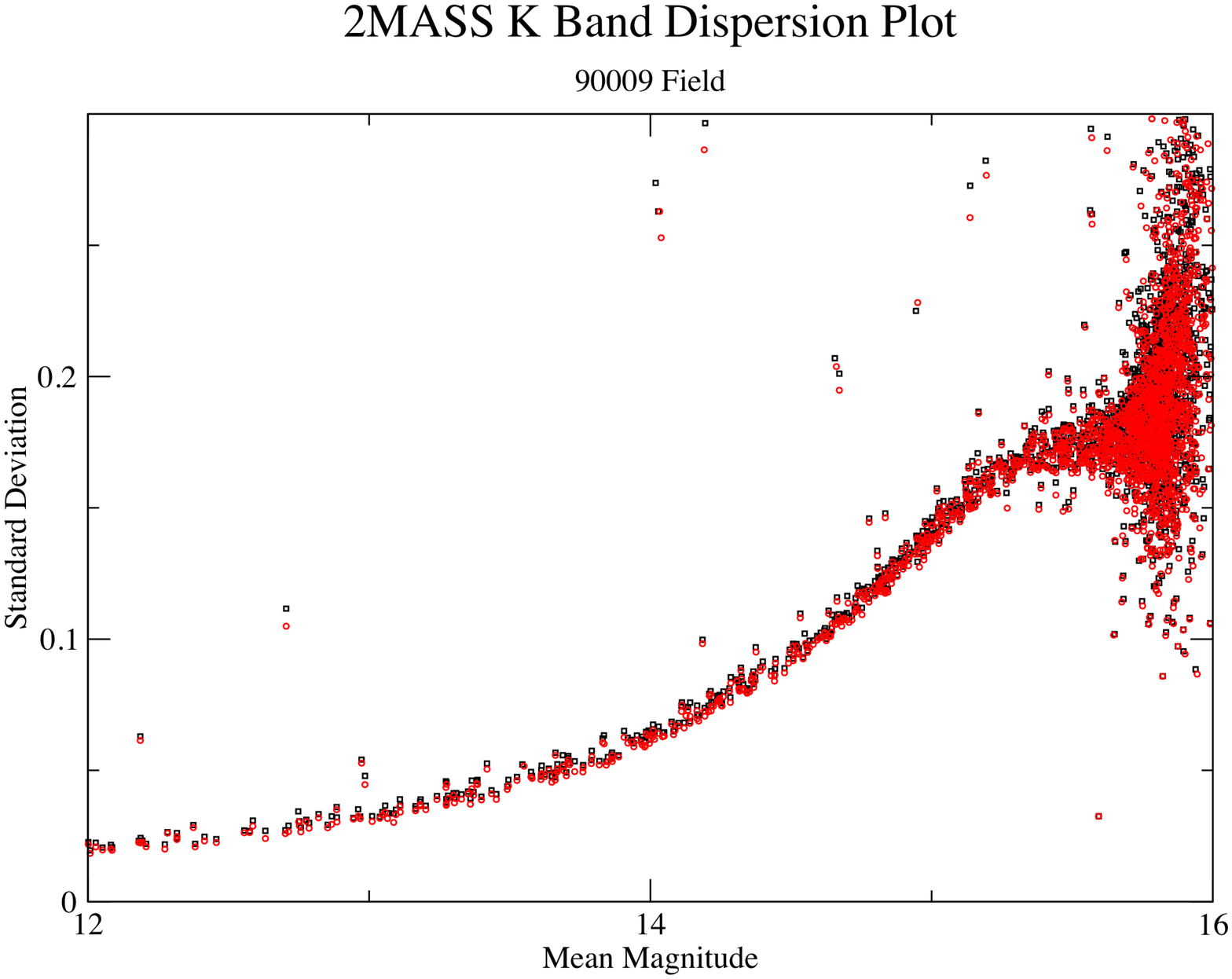}
\caption{PTF Orion pilot and 2MASS dispersion using both the original TFA and modifications. The top panel is using the unmodified version of TFA and the bottom panel is the modified version. Black points indicate light curves before detrending, and red indicates after detrending.}
\end{figure}
\begin{figure}
\epsscale{.85}
\plotone{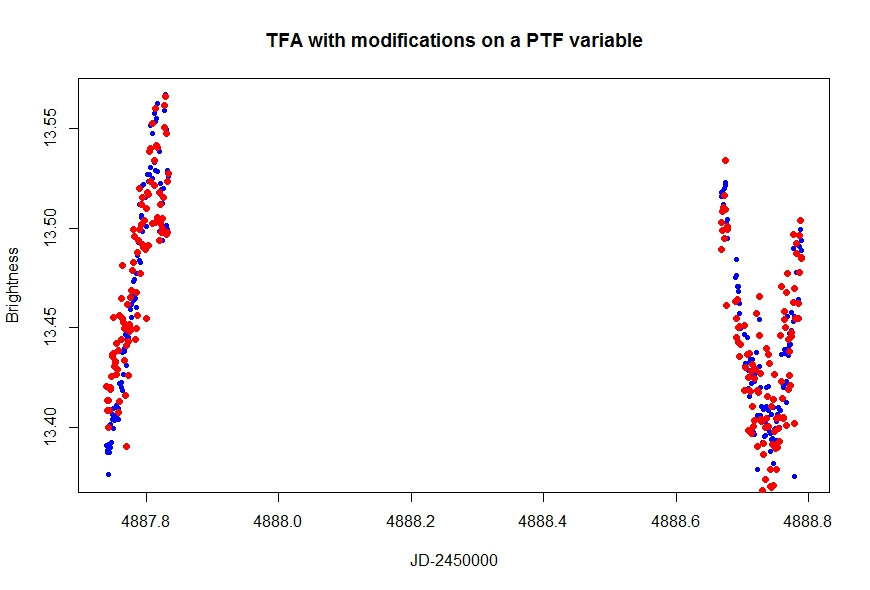}
\plotone{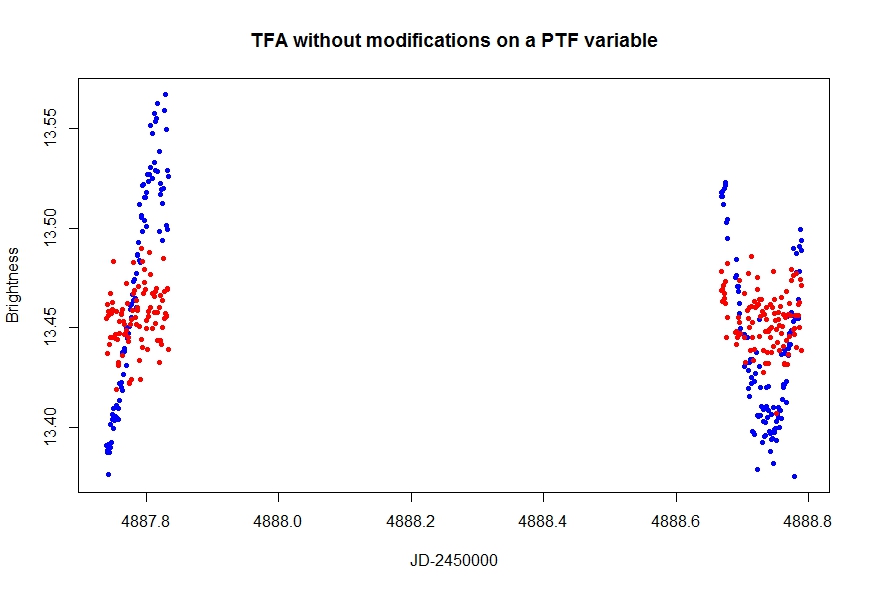}
\caption{Detrended PTF Orion pilot variable candidate using the unmodified version of TFA compared to the modified version, where the original light curve is in blue and the light curve after filtering is in red. Note that the unmodified algorithm appears to filter intrinsic variability of the light curve which is not desired. Note that TFA is used in the non-reconstructive mode}
\end{figure}
\begin{figure}
\includegraphics[scale=.65]{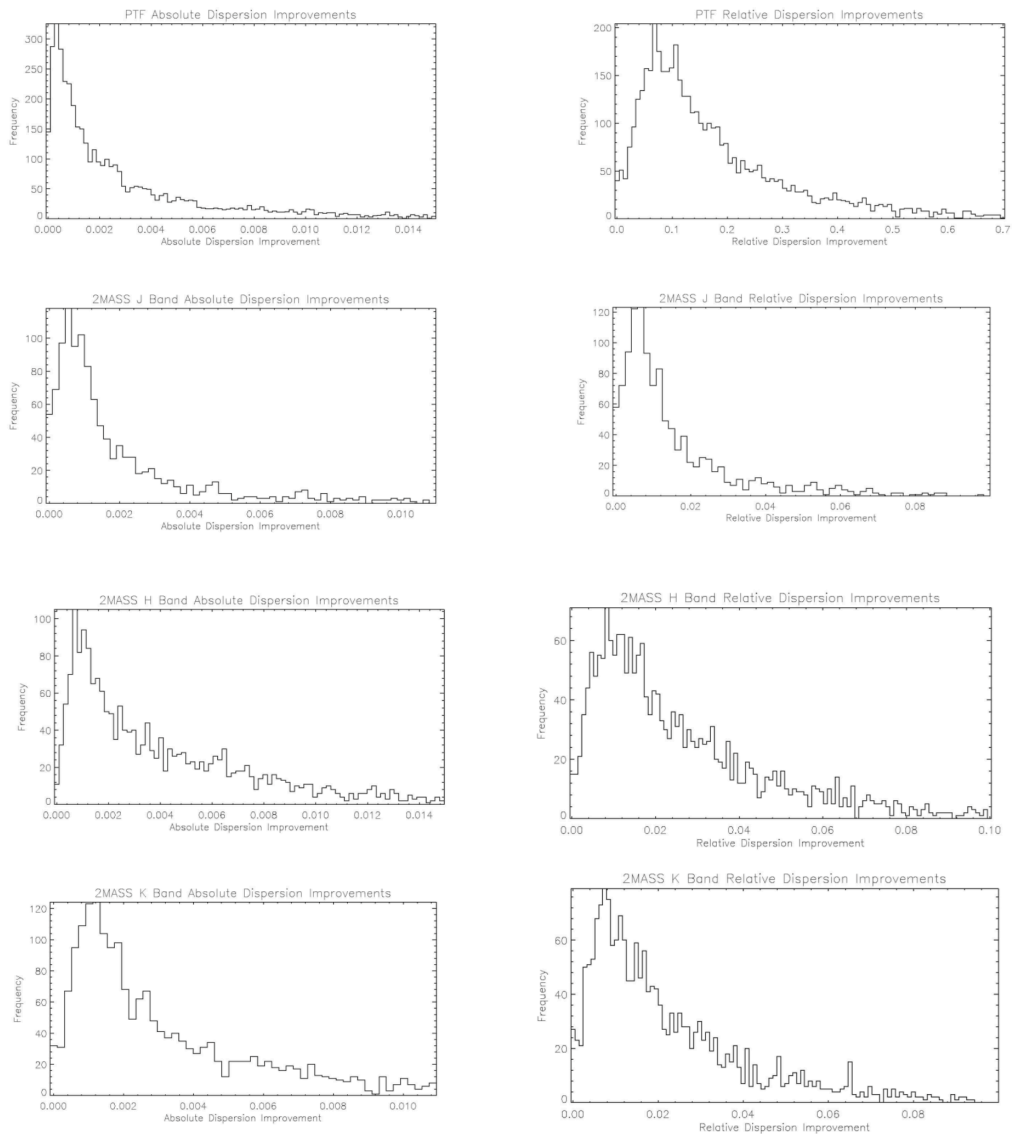}
\caption{Histograms illustrating the distribution of absolute and relative dispersion improvements using the modified version of TFA.  The left hand-side is in units of magnitudes, and the right hand side in fractional improvement in dispersion as measured in magnitudes e.g., $\frac{\sigma_{old}-\sigma_{new} }{\sigma_{old}}$}
\end{figure}

\begin{deluxetable}{crrrrrrrrrrrrr}
\tabletypesize{\scriptsize}
\rotate
\tablecaption{Summary of key parameters for synthetic light curve tests}
\tablehead{
\colhead{Parameter}&\colhead{Value}} 
\startdata
 Systematic period & 1,4,20 or 400 days \\
 Systematic amplitude & Uniformly drawn from .005, .01, or .02 mag\\
 Trasnit depth & Uniformly drawn from .005, .01, or .02 mag  \\
 Transit duration fraction & .Uniformly drawn from 005, .01, or .02\\
 Transit period & .Uniformly drawn from $\pm$ 10 $\%$ of 1, 4, 20, or 400 days \\
 Sinusoid period &  Uniformly drawn from $\pm$ 10 $\%$ of 1, 4, 20, or 400 days \\
 Sinusoid amplitude &  Uniformly drawn from .005, .01, or .02 mag \\
 Uniform measurement standard deviation & .01 mag\\
 Nightly varying measurement standard deviation & Uniformly drawn from .005 -.015 mag\\
\tableline
\enddata
\tablecomments{This table summarizes the key parameters of the synthetic light curve tests we conducted.}
\end{deluxetable}

\begin{deluxetable}{crrrrrrrrrrrrr}
\tabletypesize{\scriptsize}
\rotate
\tablecaption{Results of synthetic transit tests}
\tablehead{
\colhead{Measurement Errors}&\colhead{TFA} &\colhead{\#Template Used}  & \colhead{Uncertainty} &\colhead{Templates}&\colhead{Light Curves} &\colhead{Detections} \\
\colhead{}&\colhead{} &\colhead{}  & \colhead{Weighting} &\colhead{Selection Methodology}&\colhead{} &\colhead{} 
} 
\startdata
 Uniform & N & NA &  NA & NA & 432 & 46\\
 Nightly & N & NA &  NA & NA & 432 & 36\\
 Uniform & Y & 108 &  N & Random & 432 & 56\\
 Nightly & Y & 108 &  N & Random & 432 & 57\\
 Uniform & Y & 18 &  N & Random & 432 & 45\\
 Nightly & Y & 15 &  N & Random & 432 & 45\\
 Uniform & Y & 18 &  Y & Clustering & 432 & 47\\
 Nightly & Y & 15 &  Y & Clustering & 432 & 50\\
\tableline
\enddata
\tablecomments{This table displays the results of the synthetic transit simulation test we had conducted, where each row lists the results of each of 2 original data sets in addition to the 6 combinations we had tried, discussed in fuller detail in the simulation study description of Section 4.}
\end{deluxetable}

\begin{deluxetable}{crrrrrrrrrrrrr}
\tabletypesize{\scriptsize}
\rotate
\tablecaption{Results of synthetic sinusoid tests}
\tablehead{
\colhead{Measurement Errors}&\colhead{TFA} &\colhead{\#Template Used}  & \colhead{Uncertainty} &\colhead{Templates}&\colhead{Light Curves} &\colhead{Detections} \\
\colhead{}&\colhead{} &\colhead{}  & \colhead{Weighting} &\colhead{Selection Methodology}&\colhead{} &\colhead{} 
} 
\startdata
 Uniform & N & NA &  NA & NA & 432 & 8\\
 Nightly & N & NA &  NA & NA & 432 & 6\\
 Uniform & Y & 108 &  N & Random & 432 & 1\\
 Nightly & Y & 108 &  N & Random & 432 & 0\\
 Uniform & Y & 30 &  N & Random & 432 & 2\\
 Nightly & Y & 24 &  N & Random & 432 & 6\\
 Uniform & Y & 30 &  Y & Clustering & 432 & 3\\
 Nightly & Y & 24 &  Y & Clustering & 432 & 7\\
\tableline
\enddata
\tablecomments{This table displays the results of the synthetic sinusoid simulation test we had conducted, where each row lists the results of each of 2 original data sets in addition to the 6 combinations we had tried, discussed in fuller detail in the simulation study description of Section 4.}
\end{deluxetable}


\begin{thebibliography}

\bibitem[Aigrain et al.(2015)]{aigrain}Aigrain et al. (2015). MNRAS, 447, 2880

\bibitem[Akeson et al.(2013)]{akeson} Akeson, A.J., et al. (2013). Publications of the Astronomical Society of Pacific, Volume 125, Issue 930, pp. 989-999

\bibitem[Arthur et al.(2007)]{arthur} Arthur, D. and Vassilvitskii, S. (2007). k-means++: the advantages of careful seeding.

\bibitem[Bakos et al.(2007)]{bakos} Bakos, G., et al., (2007), APJ, 670, 826.

\bibitem[Cutri et al.(2006)]{cutri} Cutri, R.M., et al., (2006) APJ, Volume 131, Issue 2, pp. 1163-1183.

\bibitem[Foreman-Mackey et al.(2015)]{fore} Foreman-Mackey et al. 2015, Apj, 806, 215

\bibitem[Hartman et al.(2010)]{hartman} Hartman et al. (2010), MNRAS. 408, 475.

\bibitem[Kim et al.(2009)]{kim} Kim et al. (2009) MNRAS.  397 , pp. 558-568.

\bibitem[Kovacs, G. et al.(2002)]{kovacs3} Kovacs, G., Zucker, S. and Mazeh, T. (2002). A box-fitting algorithm in the search for periodic transits. A\&A 391:369-377.

\bibitem[Kovacs et al.(2005)]{kovacs1} Kovacs, G., Bakos, G., Noyes, R. (2005). A Trend Filtering Algorithm for Wide-Field Variability Surveys. MNRAS. Vol. 356, 557-567

\bibitem[Kovacs et al.(2008)]{kovacs2} Kovacs, G., Bakos, G. (2008). Application of the Trend Filtering Algorithm in the Search for Multiperiodic Signals. Comm. In Asteroseismology. Vol. 157.

\bibitem[Pal (2009)]{andras}Pal, Andras (2009). Tools for discovering and characterizing extrasolar planets.  arXiv:0906.3486v1

\bibitem[Plavchan et al.(2008)]{plavchan1} Plavchan, P., Gee, A H., Stapelfeldt, K., Becker, A.  (2008). The Peculiar Periodic YSO WL 4 in p Ophiuchus. The Astrophysical Journal, Volume 684, Issue 1, pp. L37-L40  

\bibitem[Plavchan et al.(2008)]{plavchan2} Plavchan, P., Jura, M., Kirkpatrick, J., Cutri, R., Gallagher, S. (2008) Near-Infrared Variability in the 2MASS Calibration Fields: A Search for Planetary Transit Candidates. The Astrophysical Journal Supplement Series, Volume 175, Issue 1, pp. 191-228.

\bibitem[Scargle (1982)]{scargle} Scargle, J.D. (1982) Studies in Astronomical Time Series Analysis II: Statistical Aspects of Spectral Analysis of Unevenly Spaced Data. Astrophysical Journal, 263:835-853

\bibitem[Skrutskie, M., et al.(2006)]{skrutskie}Skrutskie, M., et al. 2006, AJ, 131, 1163

\bibitem[Tamuz et al.(2005)]{tamuz} Tamuz, O., Mazeh, T..,  Zucker, S.(2005). The Sys-Rem Detrending Algorithm. MNRAS. Vol. 356, 1446. 

\bibitem[van Eyken et al.(2011)]{julian}The Astronomical Journal, Volume 142, Issue 2, article id. 60, 35 pp. (2011).

\bibitem[Wang et al.(2015)]{Wang} Wang, D., Foreman-Mackey D, Hogg, D., Schölkopf, B. Calibrating the pixel-level Kepler imaging data with a causal data-driven model arXiv:1508.01853

\bibitem[Zechmeister et al.(2009)] {zech} Zechmeister, M., Kürster, M. (2009) The Generalised Lomb-Scargle Periodogram. A new Formalism for the Floating-mean and Keplerian Periodograms. Astronomy and Astrophysics, 496:577-584 


\end{thebibliography}
\end{document}